\documentclass[conference]{IEEEtran}
\IEEEoverridecommandlockouts
\usepackage{booktabs} 

\usepackage{todonotes}

\usepackage{amssymb}
\usepackage{amsmath}
\usepackage{graphicx}

\usepackage{microtype}

\usepackage[capitalise,nameinlink]{cleveref}
\crefname{section}{Sect.}{Sect.}
\Crefname{section}{Section}{Sections}
\crefname{figure}{Fig.}{Fig.}
\Crefname{figure}{Figure}{Figures}
\crefname{table}{Tab.}{Tab.}
\Crefname{table}{Table}{Tables}
\crefname{lstlisting}{List.}{List.}
\Crefname{lstlisting}{Listing}{Listings}
\crefname{definition}{Def.}{Defs.}
\Crefname{definition}{Definition}{Definitions}

\newcommand{\eg}{e.\,g.,\ }
\newcommand{\ie}{i.\,e.,\ }

\usepackage{csquotes} 

\usepackage{listings} 

\usepackage{tikz}
\usetikzlibrary{decorations.markings}
\tikzstyle{vertex}=[circle, draw, inner sep=0pt, minimum size=6pt]

\usepackage{caption}
\usepackage{subfigure}

\usepackage{array,booktabs}

\newcommand{\labeltitle}[1]{\noindent\textbf{#1}.}

\usepackage{blindtext}
\usepackage{enumitem}

\usepackage{url}

\ifCLASSINFOpdf
\else
\fi
\begin{document}
%
\title{Catalog of Formalized Application Integration Patterns}

\author{\IEEEauthorblockN{Daniel Ritter, Stefanie Rinderle-Ma}
	\IEEEauthorblockA{University of Vienna\\
		\{firstname.lastname\}@univie.ac.at}
    \and
    \IEEEauthorblockN{Marco Montali, Andrey Rivkin}
    \IEEEauthorblockA{Free University of Bozen-Bolzano\\
        \{lastname\}@inf.unibz.it}
    \and
    \IEEEauthorblockN{Aman Sinha*\thanks{*All the work is carried out by the author is prior to joining Amazon.}}
    \IEEEauthorblockA{Amazon Research\\
    	amansinh@gmail.com}}


%


\maketitle


\begin{abstract}
    Enterprise application integration (EAI) solutions are the centrepiece of current enterprise IT architectures (\eg cloud and mobile computing, business networks), however, require the formalization of their building blocks, represented by integration patterns, verification and optimization. 
    
    This work serves as an instructive pattern formalization catalog that leads to the formalization of all currently known integration patterns.
    Therefore, we explain the classification of the underlying requirements of the pattern semantics and formalize representative patterns from the different categories, by realizing them in timed db-net.
    In this way, the catalog will allow for the addition of future patterns by assigning them to a category and applying the described formalism. 
\end{abstract}

%
\IEEEpeerreviewmaketitle

\section{Introduction}

The enterprise integration patterns (EIPs) from 2004~\cite{hohpe2004enterprise} denote messaging patterns that serve as building blocks, when implementing an enterprise application integration (EAI) system~\cite{Linthicum:2000:EAI:328930}.
While the EIPs are still practically relevant today~\cite{DBLP:journals/software/ZimmermannPHW16,Ritter201736}, the emerging technological, social and business trends since then require pattern extensions, \eg for integration adapters and endpoints~\cite{DBLP:conf/caise/0001H15}, exception handling and fault tolerance~\cite{DBLP:conf/edoc/RitterS14,ritter2016exception}, as well as information security among many other aspects~\cite{DBLP:journals/corr/RitterR15,Ritter201736}.
With the totality of those patterns playing a major role in real-world application integration architectures, the lack of a comprehensive formalization of the single pattern semantics and their compositions (beyond the currently only attempt using plain coloured petri nets (CPN)~\cite{DBLP:conf/caise/FahlandG13,fahland2012using}) will be instrumental for the verification and optimization of the current and future EAI process modeling and architectural solutions~\cite{Ritter201736}.

While we found a suitable formal representation in the recent work on db-nets~\cite{DBLP:journals/topnoc/MontaliR17} and extended them to timed db-nets for the formalization of EIPs, this work strives to collect and formalize all patterns from the aforementioned, currently known pattern catalogs, by being a catalog of formalized patterns itself.
However, with a total amount of $166$ patterns and approximately $139$ that are due to formalization (\ie no meta concepts), a complete catalog of formalized patterns would not be practical and lead to many repetitive formalizations of the same underlying concepts.

Consequently, we categorized the underlying requirements for the patterns' semantics (\ie data and control flow, external resources, transactions, complex message formats, and time) and discuss representative patterns from each of the categories (\eg data-control, data-time patterns) and their realizations in timed db-net in~\cref{sec:formalization}:
\begin{itemize}
    \item Data, transact. resource
    \item Control, transacted resource, time
    \item Control-time
    \item Data, transact. Resource Time
\end{itemize}
With these representatives of each category, the others can be easily formalized due to the same underlying semantic concepts, making this work rather an instructive catalog manual, than a simple pattern reference.
The new patterns identified after $2004$ denote a case on how to add and formalize potentially new patterns beyond this work, since they strengthened existing and added new categories.

Then in~\cref{sec:testing}, we briefly discuss how the timed db-net formalism helps to experimentally test the correctness of the patterns--again by category and not for each pattern--
and \cref{sec:conclusion} concludes this work.

\section{Formalized Patterns by Category} \label{sec:formalization}

In this section, we define selected patterns from the requirement categories discussed before.
The subsequent categories have been chosen, since the go beyond the already existing formalisms and in their combination they allow for a complete representation of the known as well as future patterns.

\subsection{Data, transact. resource patterns: Resequencer}

\begin{figure}[bt]
    \centering
    \includegraphics[width=1.0\columnwidth]{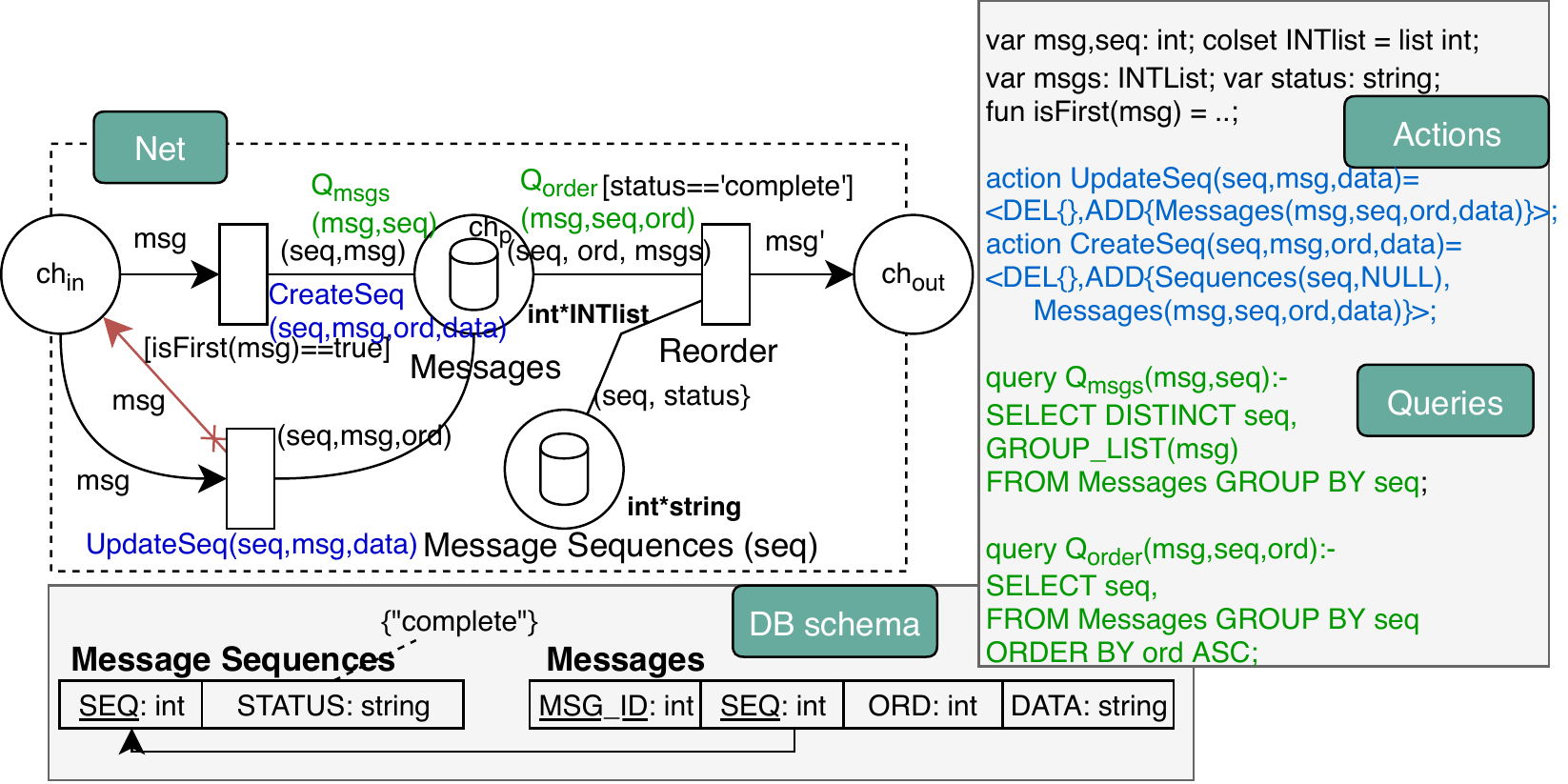}
    \caption{Resquencer pattern}
    \label{fig:resequencer}
\end{figure}

The stateful Resequencer is a pattern that guarantees the order of messages in (asynchronous) communication~\cite{hohpe2004enterprise}.
\Cref{fig:resequencer} shows the resequencer in timed db-net representation.
The entering message \texttt{msg} contains sequence (\texttt{seq}) and order (\texttt{ord}) information and is persisted in the database, represented by a db-net view place $ch_p$.
For the first message of a sequence the sequence will be created in view place \texttt{Message Sequences}, and for all subsequent message of that same sequence, the messages are stored.
As soon as the sequence is complete, \ie all messages of that sequence arrived, the messages of that sequence are queried from the database in order by the \texttt{Reorder} transition.
Eventually, the messages are forwarded in ascending order to $ch_{out}$.

\subsection{Control, transacted resource, time patterns: Circuit Breaker}

\begin{figure}[bt]
    \centering
    \includegraphics[width=1.0\columnwidth]{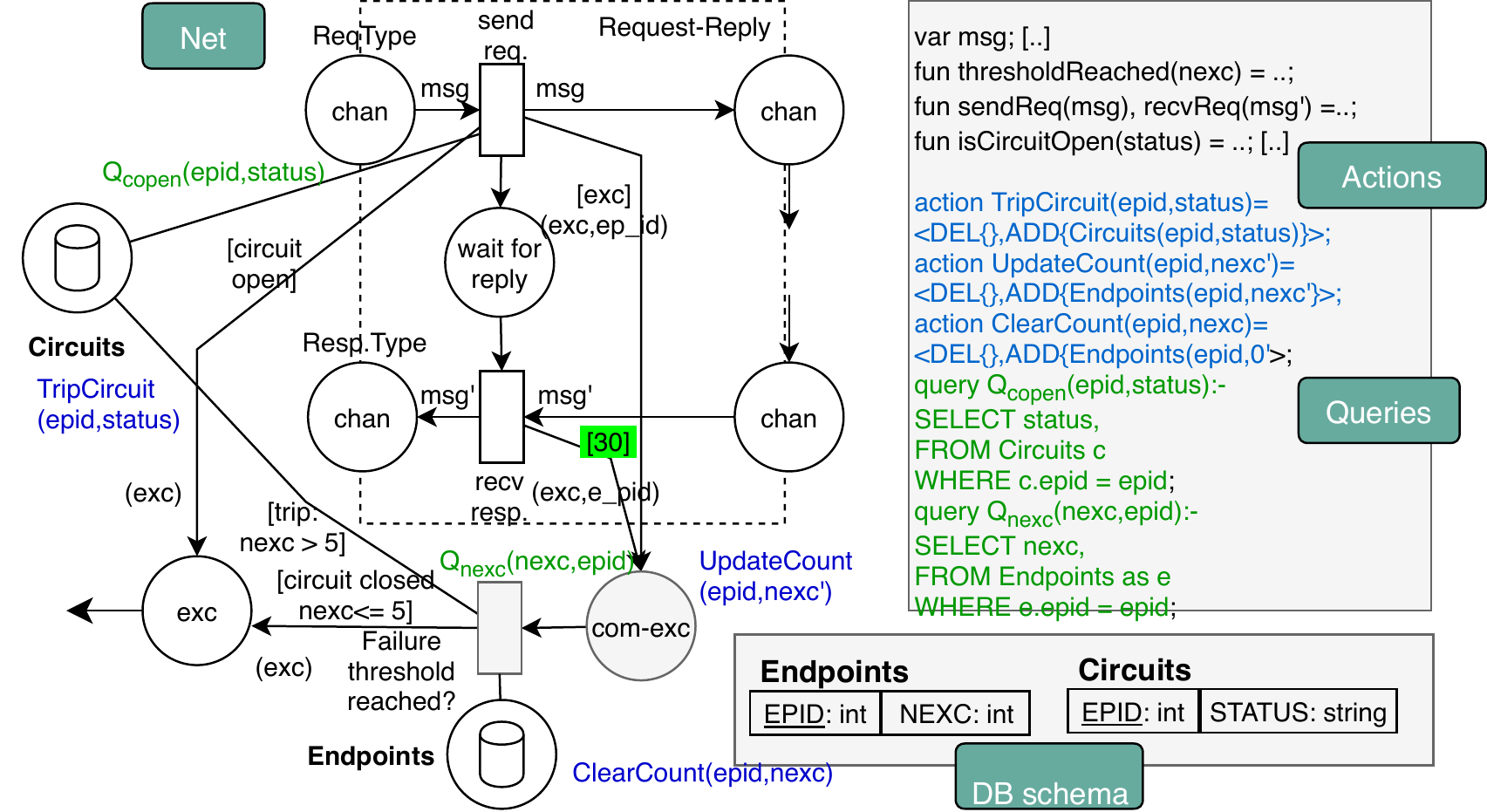}
    \caption{Circuit breaker}
    \label{fig:circuit_breaker}
\end{figure}

The Circuit Breaker pattern~\cite{Ritter201736} addresses failing or hang up remote communication, which impacts the control flow of a Request-Reply pattern~\cite{hohpe2004enterprise} by using transacted access to external resources.
\Cref{fig:circuit_breaker} shows a request-reply representation in timed db-net, extended by a circuit breaker \enquote{wrapper} (using db-net view places) that protect the remote call.

At the beginning, the circuit is closed, \ie communication is enabled.
In case of an exception (\texttt{exc}) during send or a timeout of 30 time units during receive, the number of failed attempts (\texttt{nexc}) is increased and stored into the \texttt{Enpoints} view place, which maintains a list of all endpoints (\texttt{epid}) and their consecutive failures.
If the number of failures reaches a limit (\eg nexc $>5$) the circuit trips, and thus updates the \texttt{status} entry in view place \texttt{Circuits} to \enquote{open}, which let all subsequent messages immediately go to the exception place (\texttt{exc}).
The circuit breaker can be closed by manually updating the status to \enquote{closed} to give the remote endpoint another chance.
With additional logic, self-resetting mechanisms can be implemented.

\subsection{Control-time patterns: Throttler, Delayer}

The following patterns mostly require control flow and time aspects, and are thus in timed CPNs with guards.

\begin{figure}[bt]
    \begin{center}$
        \begin{array}{cc}
        \subfigure[Throttler]{\label{fig:throttler}\includegraphics[width=0.5\columnwidth]{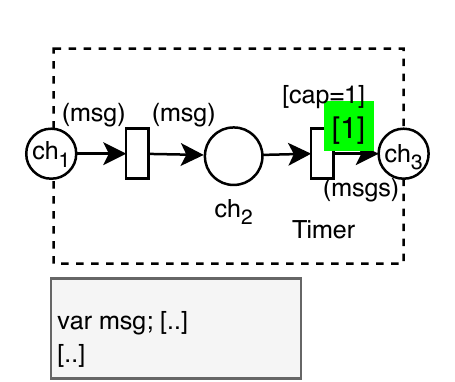}} &
        \subfigure[Delayer]{\label{fig:delayer}\includegraphics[width=0.5\columnwidth]{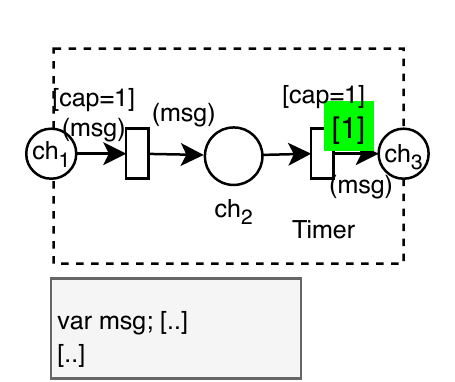}} 
        \end{array}$
    \end{center}
    \caption{Control-time patterns}
    \label{fig:}
\end{figure}

The Throttler pattern helps to ensure that a specific receiver does not get overloaded by regulating the number of transferred messages.
\Cref{fig:throttler} shows the realization of a throttler that emits five messages per second to the receiving place $ch_3$.

A slightly different pattern of this category is the delayer, as shown in~\cref{fig:delayer}, which uses a timer to reduce the frequency of messages sent to the receiving place $ch_3$.

\subsection{Data, transact. Resource Time patterns: Aggregator}

\begin{figure}[bt]
    \centering
    \includegraphics[width=1.0\columnwidth]{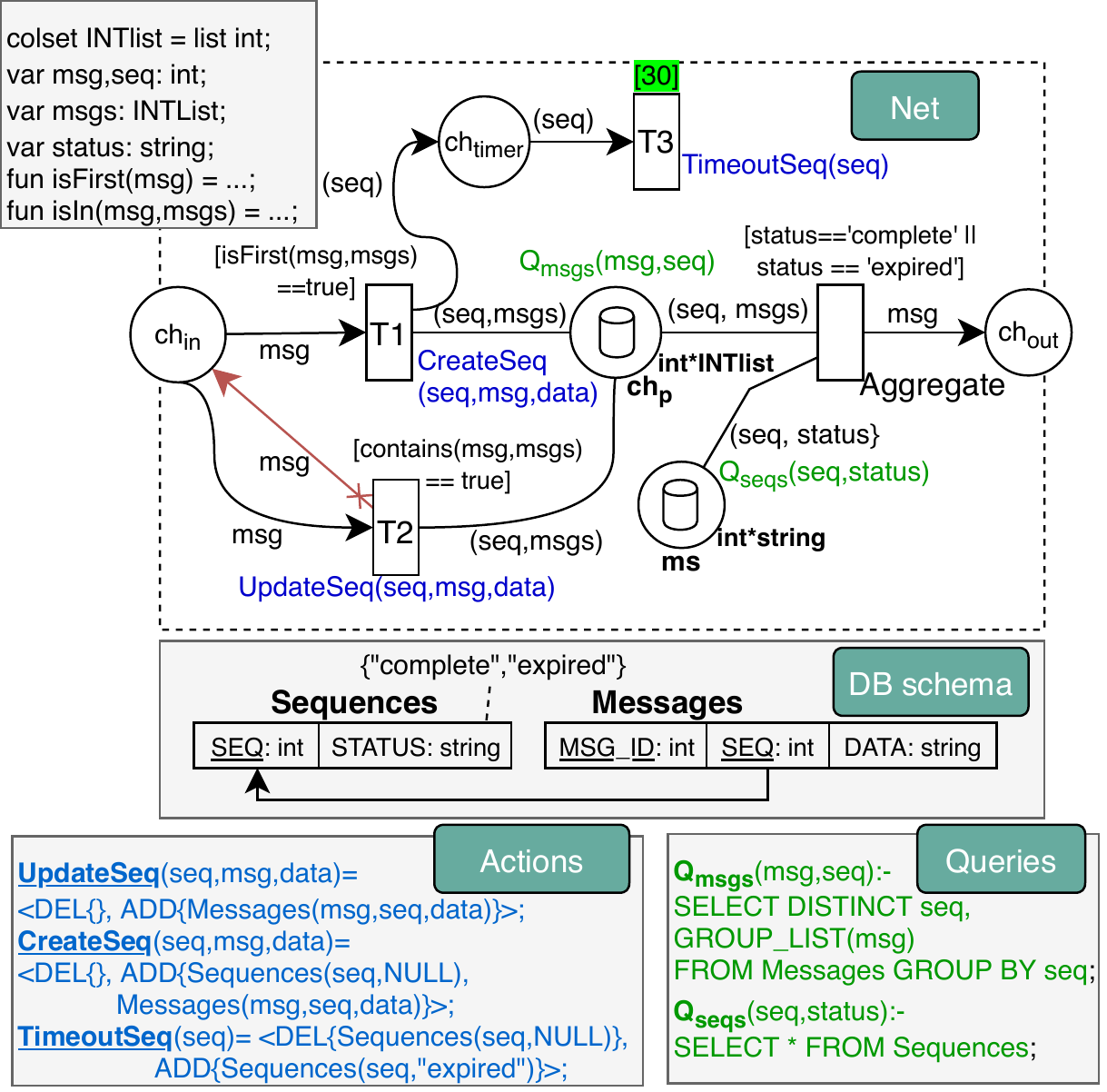}
    \caption{Aggregator pattern.} 
    \label{fig:aggregator}
\end{figure}

The combination of data, transacted resources and time aspects in patterns makes them the semantically most complex ones.
For example, \cref{fig:aggregator} specifies the semantics of a commonly used stateful Aggregator~\cite{hohpe2004enterprise} pattern.
The aggregator persistently collects messages in a special timed db-net view place $ch_p$ and aggregates them in an \texttt{Aggregate} PN transition based on a completion condition (\eg sequence \texttt{isComplete}) or on timeout, depending on a sequence \texttt{seq}, represented as PN guards.
For this an incoming message \texttt{msg} is correlated (cf. \texttt{correlate}) to an existing sequence based on its content.
If the message is the first of a sequence, a new sequence and a message to sequence assignment is created in a persistent store called \texttt{Message Sequences}.
If a message correlates to an existing sequence, which is aggregated due to a timeout \texttt{isExpired}, the update fails.
Then a roll-back is executed (red reverse timed db-net arc) that puts the message back to the message channel $ch_{in}$ (PN Place).
Now, this message matches \texttt{first msg} and a new sequence is created accordingly.

\section{Correctness Testing of timed db-net Patterns} \label{sec:testing}

\begin{figure}[bt]
    \centering
    \includegraphics[width=0.6\columnwidth]{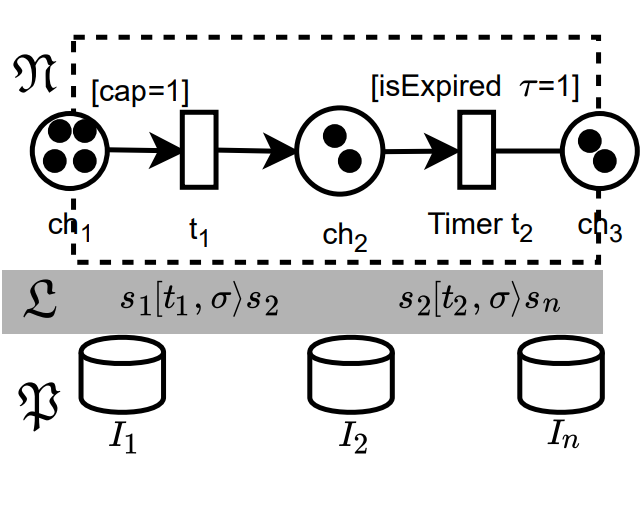}
    \caption{Throttler testing (schematic)}
    \label{fig:throttler_testing}
\end{figure}

The correctness of an integration pattern realization represented in timed db-net can be validated by evaluating the execution trace on the persistence layer.
According to the timed db-net execution semantics, a pattern produces several B-snapshots $s_1,..,s_n$ during the execution of the pattern from an input snapshot $s_1= \langle I_1,m_1 \rangle$ with database instance $s_n= \langle I_n,m_n \rangle$ to a final snapshot $s_n$, denoted by $s_1[t,\sigma \rangle s_2$,..,$s_i[t,\sigma \rangle s_n$.
In case of a database instance $I_j$ is not compliant with $\mathfrak{P}$, then the execution stops and leaves the timed db-net in an intermediate state $I_j= I_{i-1}$, otherwise $I_n$ is the final state.
Hence, the control flow can be validated, by checking, whether the pattern produces a token to the correct final database instance.
\Cref{fig:throttler_testing} gives a schematic view of an timed db-net pattern, for which the inner workings are unknown and the data is exchanged through input places $ch_1,..,ch_i$, output places $ch_{n-m},..,ch_n$, and an intermediate place $ch_j$ in $\mathfrak{N}$ that subsumes all exceptional places, together with the corresponding database intances $I_1,..,I_i$, $I_{n-m},..,I_n$, and $I_j$.
The input or newly created tokens eventually manifest in entries in the DB instances.

\labeltitle{Data and (transacted) resouce-bound patterns}
With respect to data, format 
, transacted resources
and exceptional situations, for a given instance with test data $I_1$, either an expected final persistent database instance $I_n$ with the correct schema or an expected error state $I_j$ must be produced by the pattern.
Otherwise the pattern is incorrect with respect to its definition for the requirements.

\labeltitle{Patterns with msg. channel order} Similarly, the channel execution order
can be validated.
In case of the content-based router, an initial instance $I_1$ will result a different output instance $I_n$, depending on the values in $I_1$ and the routing conditions.
While this can be checked as for the first case, the balancer requires a sequence of input instances, which then have to produce data entries in the output instances that fit the probability values and distribution of the balancer (\eg Kolmogorov-Smirnov test~\cite{1933sulla}).

\labeltitle{Time-bound patterns} Finally, a timed pattern can be validated by assigning timestamps time to the database instances (as \enquote{on-insert timestamps} in actual databases).
This allows for checking delays by comparing the insert timestamps $time(I_1)$, $time(I_n)$ of data to instance $I_1$ and those of $I_n$.
With this, a numeric delay $d$ can be checked by $time(I_n)-time(I_1)>=d$ and a message per time ratio $m/time$ by counting the number of entries \#(e) inserted to $I_n$ within time buckets $b/unit$, making pattern valid, if $\#(e)_{I_n} <=$ expected.

For instance,~\cref{fig:throttler_testing} shows a throttler pattern, provided with test messages into its input place $ch_1$ in $\mathfrak{N}$, bound to a database instance $I_1$ in $\mathfrak{P}$.
The transition $t_1$ takes the messages from $ch_1$ one after the other (\ie capacity \texttt{guard} cap=1) and inserts them into $ch_2$ with instance $I_2$, where the messages are picked up by a timed transition \texttt{Timer} and moved to $ch_3$, and thus slowing down the processing.
The data logic layer $\mathfrak{L}$ mediates by rewriting the database instances from the input $s_1[t,\sigma \rangle s_2$, to the output $s_2[t,\sigma \rangle s_3$.
Thereby the average size of $I_2$ denotes the size of the time buckets $b/unit$.


\labeltitle{Example: flawed pattern implementation}

In the absence of a tool that allows for the verification (model-checking) of a pattern, we test their correctness through simulation, by example of a \enquote{flawed} content-based router implementation.
Therefore, we use our timed db-nets CPN Tools extension\footnote{Demonstrator available for download, 10/2018: \url{https://github.com/dritter-hd/db-net-eip-patterns}; containing a flawed implementation of a content-based router, and the non-flawed implementation of the aggregator pattern from~\cref{fig:aggregator}}.

\begin{figure*}[bt]
	\centering
	\includegraphics[width=1.0\linewidth]{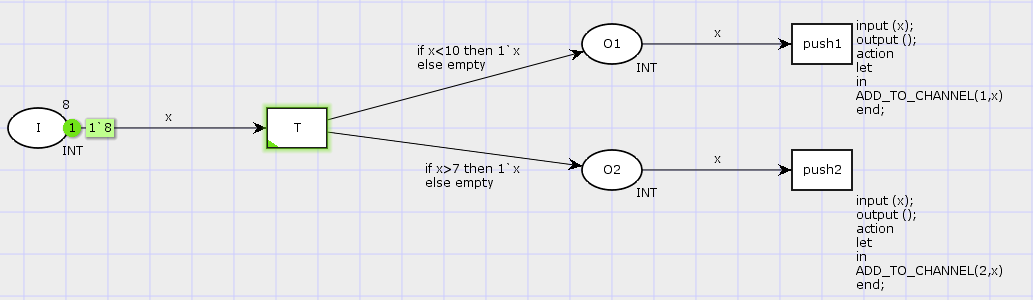}
	\caption{Flawed content-based router: transition T enabled}
	\label{fig:flawed_router_1}
\end{figure*}

A content-based router, is a pattern that takes one input message and passes it, read-only to exactly one receiver.
This is done by evaluating a condition per recipient on the content of the message.
\Cref{fig:flawed_router_1} shows one out of many router implementations, which look correct, however, violates this definition on the data and not the control level.
For the evaluation we use the aforementioned method for \enquote{data and (transacted) resource-bound patterns}, which is based on the reachability of a correct database state.
Such a correct state would be a database instance with data in table \texttt{channel1} and an empty \texttt{channel2} table.

\begin{figure*}[bt]
	\begin{center}$
		\begin{array}{cc}
		\subfigure[\texttt{channel1}]{\label{fig:channel1}\includegraphics[width=0.5\linewidth]{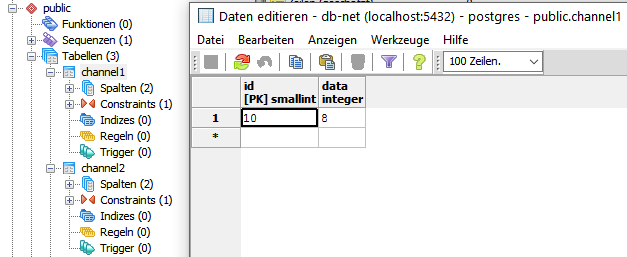}} &
		\subfigure[\texttt{channel2}]{\label{fig:channel2}\includegraphics[width=0.5\linewidth]{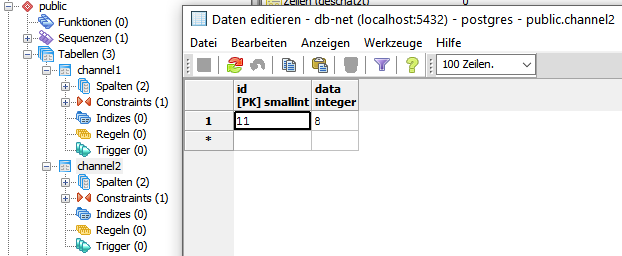}}        
		\end{array}$
	\end{center}
	\caption{Database instance after content-based router PN was executed \enquote{successfully}}
	\label{fig:}
\end{figure*}

Now, let us explore the inner workings of this implementation using timed db-net.
In~\cref{fig:flawed_router_1}, transition T reads the token in place I and then conditionally inserts it to the two subsequent places.
\begin{figure*}[bt]
	\centering
	\includegraphics[width=1.0\linewidth]{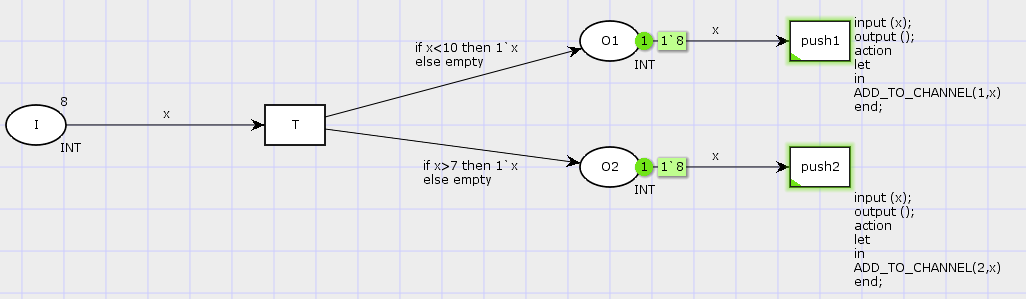}
	\caption{Flawed content-based router: messages duplicated according to the conditions / transition guards}
	\label{fig:flawed_router_2}
\end{figure*}
Since the value of the token matches all conditions, both output places $O_1$ and $O_2$ receive a copy of the token, shown in~\cref{fig:flawed_router_2}.
In terms of application integration, this could mean that two companies receive a payment request or a sales order that was actually meant for only one of them.
In the net, the two subsequent transitions $push_1$ and $push_2$ are enabled and fire by executing the database inserts $ADD\_TO\_CHANNELx$, while $x$ being the respective database table of the receiver.
From the net alone, the semantics seem to be correct.
However, on the persistence layer, no correct state has been reached.
This is illustrated by looking into the database instance after the tokens have been processed successfully on the control layer (cf.~\cref{fig:channel1}, \cref{fig:channel2}).
In the persistence layer, both tables are filled with data, which is an invalid state according to the definition of the content-based router.
Hence, the deep insight into the process and corresponding data aspects of timed db-net allow for detecting flaws in the pattern implementations as well as richer information for fixing it.

\section{Conclusion} \label{sec:conclusion}

In this work, we define and discuss the formalization of important integration patterns per category, and thus contribute an instructive catalog of pattern realizations and a description of testing their correctness for the current and future patterns to come.
The power of the chosen timed db-net formalism becomes especially clear for complex patterns, since for the first time, all of their underlying data, transactional resource and time semantics become explicit now.

\bibliographystyle{abbrv}
\bibliography{formal_v2}

\end{document}